\documentclass{emulateapj}
\usepackage[below]{placeins,epsfig}
\usepackage{natbib}
\slugcomment{to be published by the Astrophysical Journal Letters}

\def\gtorder{\mathrel{\raise.3ex\hbox{$>$}\mkern-14mu
    \lower0.6ex\hbox{$\sim$}}}
\def\ltorder{\mathrel{\raise.3ex\hbox{$<$}\mkern-14mu
    \lower0.6ex\hbox{$\sim$}}}

\def\kmsmpc{\ {\rm km~s^{-1} Mpc^{-1}}}
 
\def\hmsun{h^{-1} M_\odot}

\shorttitle{Galaxy Formation in Heavily Overdense Regions}
\shortauthors{Romano-Diaz et al.}

\begin{document}

\title{
Galaxy Formation in Heavily Overdense Regions at $z\sim 10$:\\
the Prevalence of Disks in Massive Halos}

\author{ 
Emilio Romano-D\'{\i}az\altaffilmark{1}, Jun-Hwan Choi\altaffilmark{1}, Isaac   
         Shlosman\altaffilmark{1}, Michele Trenti\altaffilmark{2}
}
\altaffiltext{1}{
Department of Physics and Astronomy, 
University of Kentucky, 
Lexington, KY 40506-0055, USA
}
\altaffiltext{2}{
Department of Astrophysical \& Planetary Sciences, CASA
University of Colorado,
Boulder, CO 80309, USA
}

\begin{abstract}
Using a high-resolution cosmological numerical simulation, we have analyzed the 
evolution of galaxies at $z\sim 10$ in a highly overdense region of the universe. 
These objects could represent the high redshift galaxies recently observed by the 
Hubble's WFC3, and be as well possible precursors of QSOs at $z\sim 6-7$. To 
overcome the sampling and resolution problems in cosmological simulations of these
rare regions, we have used the Constrained Realizations method. Our main 
result for $z\sim 10$ shows the high-resolution central region of 
$3.5h^{-1}\,{\rm Mpc}$ radius in comoving 
coordinates being completely dominated by disk galaxies in the total mass range of 
$\gtorder 10^9h^{-1}\,{\rm M_\odot}$. We have verified that the gaseous and stellar 
disks we identify are robust morphological features, capable of surviving the 
ongoing merger process at these redshifts. Below this mass range, we find a sharp 
decline in the disk fraction to negligible numbers. At this redshift, the disks 
appear to be gas-rich compared to $z=0$, and the
dark matter halos baryon-rich, by a factor of $\sim 2-3$ above the average fraction of 
baryons in the universe. The dominance of disk galaxies in the high density peaks 
during the epoch of reionization is contrary to the 
morphology-density trend observed at low redshifts.
\end{abstract}

    
\section{Introduction}
\label{sec:intro}

Deep sky surveys are providing large sample of color-selected galaxies out 
to $z\sim 6$ (e.g., Bouwens et al. 2007) and a growing number of spectroscopically 
confirmed objects (e.g., Fan et al. 2004; Malhotra et al. 2005; Stark et al. 2011). 
With the availability of the Wide Field Camera 3 
(WFC3) on the Hubble Space Telescope (HST) more than 100 galaxies have been detected 
at $z\sim 6-8.5$ (e.g., Bouwens et al. 2010; McLure et al. 2010; Oesch et al. 2010).
These high-$z$ galaxies are expected to be significantly clustered in the high density 
environments (e.g., Overzier et al. 2006). 

Recent studies of high-redshift QSOs have indicated that they reside in rare 
and highly overdense regions at $z\sim 6$, with the comoving space density of $\sim 
(2.2\pm 0.73)\,h^3\,{\rm Gpc^{-3}}$ (e.g., Fan et al. 2004). These dense environments, 
inhabiting $\sim 1$~Gyr old Universe, can be precursors of 
massive clusters of galaxies at $z=0$ (e.g., Fan et al. 2004; Springel et al.
2005; Li et al. 2007), but not necessarily the most massive clusters 
(e.g., Trenti et al. 2008; Romano-Diaz et al. 2011). Chemical composition inferred
from QSO spectra and additional correlations with the cosmic star formation history
indicate a possible link between galaxy evolution and the QSOs
--- being metal-rich they plausibly are located at the centers of massive galaxies
(Graham et al. 2001; Barth et al. 2003; Bunker et al. 2004). 

{\it Galaxy evolution in these overdensities can differ
from that in the field}, both quantitatively and qualitatively, due to earlier formation
times, an absence or a decreased intensity of the background UV field, and a 
possible radiative and mechanical feedback from the QSOs. The reionization of 
the Universe could start with these regions as early as $z\sim 15$ and extend to 
$z\sim 6$ (e.g., Becker et al. 2001; Barkana 2002; Cen \& McDonald 2002). But the 
population of galaxies responsible for re-ionization of the Universe is not yet found.
It is also not clear at present what fraction of overdense regions and to what extent is
affected by QSOs and background UV. 

The most overdense peaks are hard to sample with cosmological simulations at high 
resolution. Recent detections of candidate galaxies and proto-clusters at $z\sim 6-10$, 
have inspired us to study properties of galaxies at these redshifts (e.g., Trenti et al.
2011a,b; Bouwens et al. 2011; Oesch et al. 2011; Capak et al. 2011), and this is
first in the series of papers on this issue.
Circumventing the difficulty of sampling, we resort to the Constrained
Realizations (CR) method (e.g., Hoffman \& Ribak 1991), a fully
self-consistent method which allows to simulate such rare regions at highest resolution,
without loss of generality.

Current efforts in modeling galaxy evolution aim at reproducing the observed morphology
at $z=0$ (e.g., Governato et al. 2004; Robertson et al. 2004; Tasker \& Bryan 2008),
but also analyze the relevant physical processes at high $z$ (e.g., Gnedin et al. 2009;
Romano-Diaz et al. 2009; Gnedin \& Kravtsov 2010; Greif et al. 2010; Wise et al. 2010; 
Pawlik et al. 2011).

To probe the dark matter (DM) backbone of galaxy formation at $z\sim 6-10$ in highly 
overdense regions, Romano-Diaz et al. (2011) have performed a set of carefully designed DM 
simulations with CRs.  
Constructed halo Mass Functions have shown that on average, the QSO hosts have 
enhanced the halo formation in their vicinity due to pure gravitational effects, when 
compared with an average region in the Universe. Here we extend our study to include 
treatment of the baryonic component,
and analyze the properties of galaxies at $z\sim 10$. In this Letter, we focus on the
most general properties of galaxy population, such as morphology and baryon
fraction. In the long run, our goal is to quantify the environment of high-$z$ 
QSOs, understanding the
similarities and peculiarities of galaxy evolution in the overdensest regions at 
$z\sim 6-10$. While current simulation does not include the QSO feedback, and for 
$z\sim 10$ neglects the UV background, it would be interesting to account for these 
effects in the future.

This Letter is structured as follows: section~2 summarizes the technical
details of our simulation setup and its initial conditions; section~3 
analyzes obtained galaxy morphology at $z\sim 10$,  and the last section 
discusses the obtained results.

\section{Numerics and Initial Conditions}

We use the modified version of the tree-particle-mesh Smoothed Particle Hydrodynamics 
(SPH) code GADGET-3 originally described in Springel (2005), in its conservative entropy 
formulation (Springel \& Hernquist 2002). Our conventional code includes radiative 
cooling by H, He, and metals (Choi \& Nagamine 2009), 
star formation, its feedback, a phenomenological model for galactic winds, and 
sub-resolution model of multiphase interstellar medium, ISM (Springel \& Hernquist 2003). 
Since we only focus on high redshift before the full reionization and do not implement 
on-the-fly radiative transfer of ionizing photons, the UV background is not included in our 
simulation. In the multiphase ISM model, star forming SPH particles contains the cold phase 
that forms stars and the hot phase that results from a supernova (SN) heating. The cold 
phase contributes to the gas mass, and the hot phase contributes to gas pressure.
For the star formation, we use the ``Pressure model'' which reduces the high-$z$ 
star formation rate (Choi \& Nagamine 2010) relative to the previous model by Springel
\& Hernquist (2003). The star formation is triggered when the gas density is above the 
threshold $n_{\rm H,SF} = 0.6\,{\rm cm^{-3}}$.

The initial conditions have been generated using the CR method 
(Hoffman \& Ribak 1991; see also Romano-Diaz et al. 2007) and are similar to those used by 
Romano-Diaz et al. (2011, see the Appendix there). The constraints were imposed onto a 
grid of $1,024^3$ within cubic box of $20h^{-1}\,{\rm Mpc}$ to create a DM halo seed 
of $10^{12}\,\hmsun$ collapsing by $z\sim 6$, according to the top-hat model. We assume 
the $\Lambda$CDM cosmology with WMAP5 parameters (Dunkley et al. 2009), 
$\Omega_{\rm m}=0.28$, 
$\Omega_\Lambda=0.72$, $\Omega_{\rm b}=0.045$, and $h=0.701$, where $h$ is the Hubble 
constant in units of 
$100\,\kmsmpc$. The variance $\sigma_8=0.817$ of the density field convolved with the 
top hat window of radius $8h^{-1}$\,Mpc$^{-1}$ was used to normalize the power spectrum.

The evolution has been followed from $z=199$. The growth of structure has been 
substantially accelerated by resulting high overdensity, $\sim 20$, within the central region.
The simulation runs in comoving coordinates and with vacuum boundary conditions. 
We have applied the zoom-in technique with 3 levels of refinement to increase the numerical 
resolution. The inner region has a radius $3.5h^{-1}\,{\rm Mpc}$, having an effective 
resolution of $1,024^3$ in DM and SPH particles. Within this region, we obtain the 
particle mass of $4.66\times 10^5\,{\rm M_\odot}$ (DM), $1.11\times 10^5\,{\rm M_\odot}$ 
(gas) and $5.55\times 10^4\,{\rm M_\odot}$ (stars). Gravitational softening is 
$\epsilon_{\rm grav}= 300$~pc (comoving).

We have used the halo finding algorithm HOP (Eisenstein \& Hut 1998) to identify  
DM halos, and have defined halo virial masses and radii in the context of the 
spherical top-hat collapse model, $M_{\rm vir} = 4/3\pi \Delta(z)\rho(z) R_{\rm vir}^3$, 
where $\rho$ is the cosmic background density and $\Delta(z)$ (Bryan 
\& Norman 1998) is the critical overdensity at virialization.  
The halo and disk shapes have been found using the method
described in Romano-Diaz et al. (2009) following Heller et al. (2007). 

\figurenum{1}
\begin{figure}[ht!!!!!!!!]
\vbox to2.8in{\rule{0pt}{2.2in}}
\includegraphics{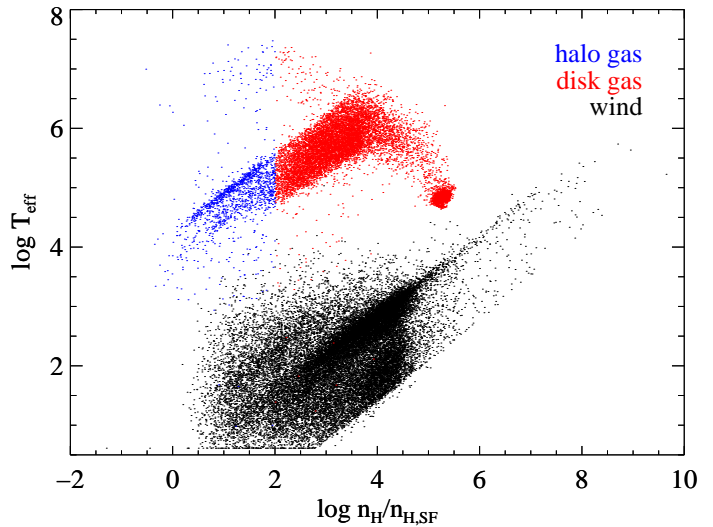}
\caption{Example of the phase diagram $T_{\rm eff}-n$, effective gas temperature vs
density, for the gas within the most massive DM halo at $z=10.2$. Density is
normalized by $n_{\rm H,SF} = 0.6\,{\rm cm^{-3}}$ --- the critical 
density for the onset of star formation (see text). The colors are: disk gas particles 
(red), halo gas particles (blue), and wind particles (black). The blue-red boundary
represents the empirically determined transition between the disk and halo gas (see text). 
This snapshot corresponds to a post-merger evolution with a spike in star formation rate.}
\end{figure}

\figurenum{2}
\begin{figure*}[ht!!!!]
\begin{center}
\includegraphics[angle=0,scale=0.75]{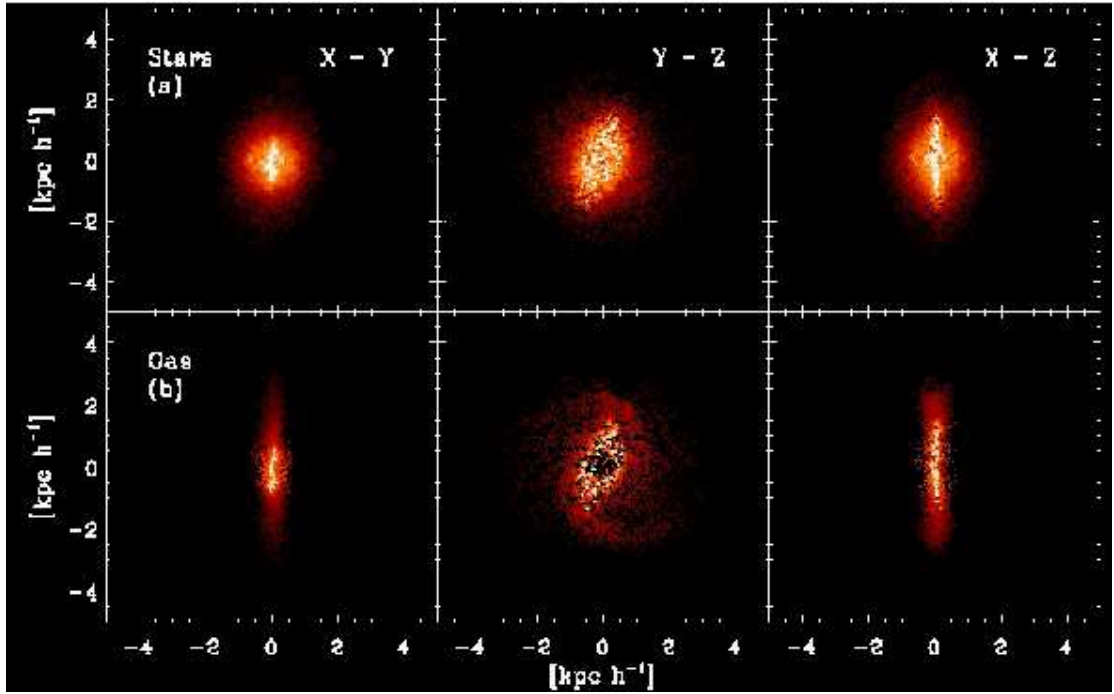}
\end{center}
\caption{Example of a deprojected face- and edge-on strongly barred disk embedded in the 
most massive halo at $z\sim 10.2$, stars (top) and gas (bottom). The frames are
$5\times 5\,h^{-1}\,{\rm kpc}$. Left frames: bar along the line-of-sight; 
right frames: bar normal to the line-of-sight. 
The total mass is $M_{\rm tot}\sim 1.11\times 10^{10}h^{-1}\,{\rm M_\odot}$,
the total disk mass is $\sim 2.9\times 10^9h^{-1}\,{\rm M_\odot}$, The gas disk shown 
corresponds to the red colored zone in Fig.~1.
}
\end{figure*}
\figurenum{3}
\begin{figure}[h!!!!!!!!!!!!!!!!!!!]
\vbox to4.8in{\rule{0pt}{2.2in}}
\includegraphics{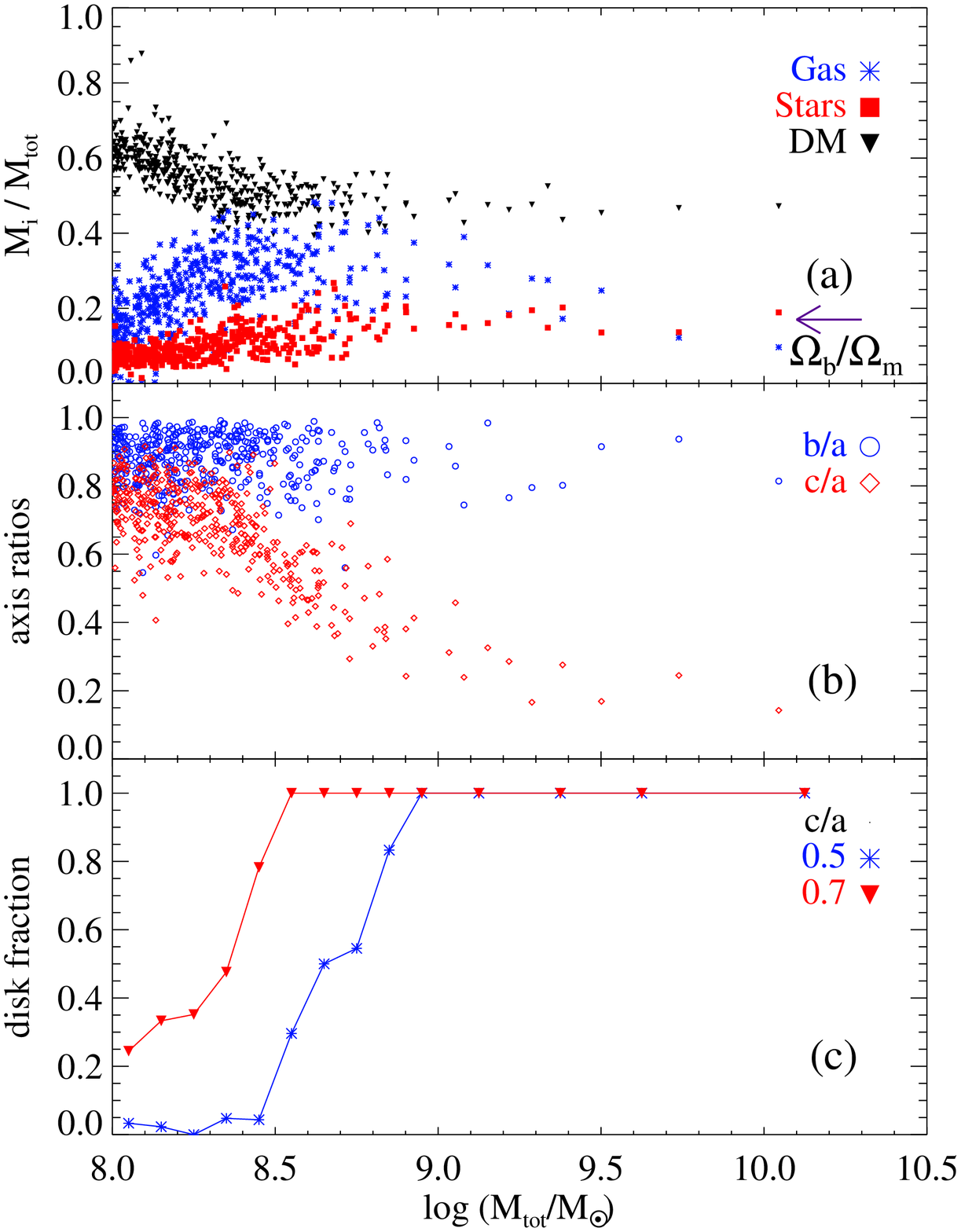}
\caption{Properties of halos and galaxies at $z\sim 10.2$ within the high-resolution 
region of $3.5h^{-1}$\,Mpc in radius, with binning of 0.25 in 
log\,$(M_{\rm tot}\,M_\odot) \gtorder 9.0$ and 0.1 dex for smaller masses, in $2\times 1,024^3$
simulation. {\it (a)} disk gas, 
disk stars and DM fractions within halos as a function of the halo total mass (species: DM --- 
black, stars --- red, gas --- blue). The arrow shows the WMAP baryon fraction, $\Omega_{\rm
b}/\Omega_{\rm m}$; {\it (b)} gas disks axial ratios $b/a$ (blue) and $c/a$ (red); 
{\it (c)} disk fraction, $f_{\rm d}$, as a function of the halo total mass, $M_{\rm tot}$. 
Shown are two defining criteria for the gas disk, $c/a\ltorder 0.5$ (blue) and $c/a\ltorder 
0.7$ (red). Halos in the process of merging at this $z$ have been omitted, overall 4
objects within this mass range --- all of them had more than one disk per halo. The control
$2\times 512^3$ simulation gave nearly identical results for Fig.~3c.
}
\end{figure}

\section{Disk Fraction}

By $z\sim 10$, nearly 500 halos above $10^8h^{-1}\,{\rm M_\odot}$ have been detected and 
cataloged. To analyze the baryons within the DM halos, they have been divided into three 
groups: (1) gas; (2) stars; and (3) gas which currently 
experiences feedback from stellar evolution and is found on ballistic trajectories --- 
this is the wind. To identify and quantify gas properties, we have 
constructed the $T_{\rm eff}-n$ phase diagram (e.g., Fig.~1). 
The gas tends to cluster in two main regions: the top ones (blue and red, see below) are 
the disk/halo particles which can experience the star formation and whose
effective temperature, $T_{\rm eff}$, is the result of a shock heating by the 
SN. The second region (black) consists of particles  
that comprise the wind. All gas can be cooled/heated via expansion/compression.

We do not know  {\it apriori} whether a halo contains a gas disk, and, therefore, 
identify its presence, based on axial ratios determined from the moments of inertia
(Heller et al. 2007; Romano-Diaz et al. 2009). Major and minor equatorial 
axes are defined as $a$ and $b$, and $c$ (disk semi-thickness) along rotation axis of 
the disk. Specifically, we focus on the ratio of the minor-to-major axis, $c/a$ --- 
its actual values are discussed below. If, based on the $c/a$ ratio, we decide that a 
particular halo does not contain a gas disk, we call the gaseous entity irregular.
We have empirically determined that disk and halo gas can be separated by a density cut 
at $n/n_{\rm H,SF} \sim 100$. 
The red color in Fig.~1 corresponds to the disk gas and the blue color to the halo gas.
We have verified this separation randomly for a large sample of DM halos.
Of course, in the absence of a disk this separation is not important, but we still
keep it for the sake of simplicity. The disk stars have been separated from halo
stars based on the alignment of their angular momenta (e.g., Romano-Diaz et al. 2009).

After isolating the disk gas, we have deprojected its spatial
distribution in each halo onto three fundamental planes: face-on and edge-on (see an
example in Fig.~2 for the most massive halo). The
disks are clearly visible, have sizes $R_{\rm d }\ltorder 3h^{-1}$\,kpc and are embedded in 
halos of $R_{\rm vir}\ltorder 7h^{-1}$\,kpc. The face-on disks show large-scale bars,
sometimes comparable to the disk size, and their gas layer full thickness is 
$\sim 1-2h^{-1}$\,kpc, and is pressure-supported. 

Figure~3a shows the disk gas and stars, and DM fractions as functions of the total 
halo masses. The cutoff at log\,$M_{\rm tot}\sim 8.0$ approximately corresponds to 
$\sim 350$ particles within $R_{\rm vir}$ and we consider this to be reliable in
determining halo (e.g., Trenti et al. 2010) and disk shapes. Two types of 
behavior can be observed in Fig.~3a. The DM fraction drops
and gas fraction increases to ${\rm log}\,M_{\rm tot}\sim 8.5$, and the
stellar fraction increases up to ${\rm log}\,M_{\rm tot}\sim 9.5$. The DM and stellar
fractions flatten thereafter, while the gas fraction declines.

The stellar disks are generally less massive than the gaseous 
ones for most of the mass range,  log\,$M_{\rm tot}\ltorder 9.5$ (Figs.~2b, 3a).
The overall trend is a strong increase in the gas fraction {\it in disks} with the 
total mass, up to log\,$M_{\rm tot}\sim 8.5$, to 
$\sim 1/3$, and a decrease thereafter. The steep decline in the gas fraction below 
log\,$M_{\rm tot}\sim 8.5$ is related to the inability of the halo to retain all
of the gas because of the SN feedback and low virial temperatures.  
Below log\,$M_{\rm tot}\sim 8.0$ (not shown here), the gas 
fraction in the halos is negligible because of the same effect. For the
most massive halos, there appears to be roughly equal fraction of disk gas and stars, 
$\sim 50\%$.  

The shapes of the gaseous disks show interesting trends (Fig.~3b). Overall 
disks appear mildly non-axisymmetric, $0.8\ltorder b/a\ltorder 1$, with a slight 
increase in the dispersion of $b/a$ toward smaller $M_{\rm tot}$. On the other hand,
there appears to be a strong trend toward increased flatness, $c/a$, with increasing
mass, down to $c/a\sim 0.1$. Below ${\rm log}\,M_{\rm tot}\sim 8.3$, the
gaseous disks appear prolate and their rotation axis can be distinguished only based 
on the angular momentum. We also note that parent halos of galaxies discussed 
here are quite triaxial with the range of $b/a\sim 0.6-1$ and $c/a\sim 0.2-0.8$ and
therefore can influence the shape of the disk they harbor (Berentzen \& Shlosman 
2006).

The stellar disks (e.g., Fig.~2b) exhibit a high density midplane component of newly 
formed stars, with the mass of $\sim 8\times 10^8h^{-1}\,{\rm M_\odot}$ for the most 
massive halo at $z=10.2$, and a lower density spheroidal stellar component, of 
$\sim 1.2\times 10^9h^{-1}\,{\rm M_\odot}$, when integrated over 
the extended halo. Note that {\it the star formation we measure is
not limited to the disk regions only, but engulfs the entire halos} because the gas 
density there appears above the threshold density $n_{\rm H,SF}$. 
So the spheroidal component in our galaxies has its origin in the halo and was not
formed as a result of the secular or dynamical thickening of the stellar disk, or 
from the heating by minor mergers. It is not clear what is the fate of this spheroidal 
stellar component. Does it represents low metallicity halo stars at low redshifts?
Following the evolution of the relative masses of disk and halo stars, we only comment
that the spheroids have formed early, by $z\sim 12$, and we observe a steadily increasing
disk masses, compared to the halo stars, with time.

We are now in a position to estimate the disk fraction, $f_{\rm d}$, within the high 
resolution box
of the simulation. To show the robustness of our result, we adopt two definitions
of the gaseous disks, these of $c/a \ltorder 0.5$ and $c/a\ltorder 
0.7$. The first and more conservative definition exhibits a sharp decline below
${\rm log}\,M_{\rm tot}\sim 9$, while the second one shows a very similar decline
below ${\rm log}\,M_{\rm tot}\sim 8.6$. Hence, both agree that the disk fraction is
100\% for the more massive halos (Fig.~3c). There is some indication that the disk
fraction levels off again for very small halos, at the level of $f_{\rm d} 
\sim 0.05-0.2$. 

We have checked the sensitivity of $f_{\rm d}$ to our definition of the disk gas 
threshold (\S3). Varying the threshold by $\pm 10\%$ does not change $f_{\rm d}$ for 
${\rm log}\,M_{\rm tot}\gtorder 9$, and changes it negligibly for smaller masses.
Furthermore, we have tested $f_{\rm d}$ in lower resolution $2\times 512^3$ simulation 
and find it nearly identical ---a sign of numerical convergence. Finally, we note that 
the blue and red point
scatter is uniform across the mass range in Fig.~3b --- a sign that the resolution
effects do not compromise the results.
In summary, the disk fraction is $f_{\rm d}\sim 1$ for for massive halos in our 
simulation, and exhibits a sharp decline by an order of magnitude
below ${\rm log}\,M_{\rm tot}\sim 8.5-9.0$.  

The circular velocities for massive halos are contributed by baryons
and DM nearly equally. Only for the most massive halo, we see maximal $v_{\rm c} 
\gtorder 100\,{\rm km\,s^{-1}}$. For the 15 most massive halos, we have instead maximal
$v_{\rm c} \gtorder 50\,{\rm km\,s^{-1}}$. Overall, the total mass in a halo roughly 
correlates with the maximal $v_{\rm c}$, meaning that the halo concentration is
a monotonic function of $M_{\rm tot}$. 

\section{Discussion}

Using a high-resolution cosmological numerical simulation, we have analyzed the 
evolution of galaxies at $z\sim 10$ in a highly overdense region
of the universe. These objects could represent high redshift galaxies recently 
observed by the WFC3/HST, and be possible precursors of QSOs at $z\sim 6-7$. 
For this purpose 
we have employed high-resolution simulations of $3.5h^{-1}\,{\rm Mpc}$ region
in radius involving a multiphase ISM and feedback from stellar evolution.
Halos with $M_{\rm tot}\sim 10^8h^{-1}\,{\rm M_\odot}$ are resolved with $\sim 350$ 
particles. 
We find that the fraction of disk galaxies is unity for total masses above
$\sim 10^9h^{-1}\,{\rm M_\odot}$, with a sharp decline in $f_{\rm d}$ for less
massive halos. We also find that the baryon fraction (with respect to total
mass) peaks around ${\rm log}\,M_{\rm tot}\sim 8.5$, slowly declines towards higher masses
and sharply declines toward smaller masses. So overall, above the resolved total
mass of $10^8h^{-1}\,{\rm M_\odot}$, halos are baryon-rich at these redshifts 
with respect to the WMAP $\Omega_{\rm b}/\Omega_{\rm m}$ ratio, in 
a broad agreement with Harford et al. (2008). More
massive halos appear with about equal gas and stellar fractions, while less massive
ones have higher gas fractions. 

The prevalence of disk galaxies at high redshifts is not a trivial result because
of ongoing merging between galaxies which contributes to strong perturbations in
disk morphology, and has been tracked by us. Four objects in the process of merging
have been omitted from our statistics for $f_{\rm d}$. All the 
mergers we observe are between disks. Despite this process, we find that
gas-rich halos are capable of reforming the disks in a short time. This effect
has been demonstrated previously by Springel \& Hernquist (2005) and we confirm it 
here --- the mergers we have observed do not destroy the disks. Taken at 
the face value, the dominant population of disk galaxies at overdense peaks at 
$z\sim 10$ appears to contradict the morphology-density relation observed at
low $z$ (e.g., Dressler 1980). The most straightforward explanation for this is
that various processes leading to deficiency of disks in the density peaks at
$z=0$ (e.g., dry mergers, gas ablation) have had not enough time to operate at $z\sim 
10$. A number of processes at high $z$ contribute to galaxies being gas-rich,
among them the ongoing gas accretion which joins the disks in their equatorial planes. 
In tandem with prominent (gas-rich) bars, these disks can serve as a fuel reservoir 
for growing supermassive black holes (SBHs) already at this epoch.

While it has been claimed that the feedback from such SBHs (neglected here) can 
terminate the accretion
process and drive the gas out of the galaxy (e.g., Di Matteo et al. 2005), it is
by no means a solved problem. It is also not clear when the morphology-density
correlation observed here is reversed to its low-redshift counterpart. 

Our simulation has demonstrated that well resolved
galaxies hosted by ${\rm log}\,M_{\rm tot}\gtorder 9$ mass halos possess well
defined gaseous and stellar disks. Moreover, by observing the evolution of these
galaxies through major mergers, we have verified that the disks are also robust
morphological features and survive this process. We do agree with Greif et al. 
(2010) and Wise et al. (2010)
simulations of smaller halos that the stellar feedback becomes increasingly
dominant when smaller, $\sim 10^8h^{-1}\,{\rm M_\odot}$ halos are analyzed, and with
Pawlik et al. (2011) who have obtained the disk formation in higher mass halos. 
We interpret the increased irregular morphology below 
${\rm log}\,M_{\rm tot}\sim 9.0$ as the effect of a stellar (SN) feedback on the disk
gas. At this mass range, the morphology becomes chaotic and the gas fraction declines
sharply. The observed decreasing disk fraction in low mass halos appears to be real. 
However, the analysis of the feedback process to quantify this phenomenon is outside 
the scope of this Letter.

The dominant population of the disk galaxies at $z\sim 10$ in the overdense regions 
simulated in this work appear consistent with the presence of extended morphologies in 
bright $z\sim 7$ galaxies observed by HST (Oesch et al. 2010). However, a direct 
comparison of our predictions will be possible only with next generation telescopes such 
as JWST or 30m-class observatories from the ground, which will have sufficient sensitivity 
and angular resolution to investigate the morphology of $z\gtorder 8$ galaxies.

\acknowledgements
We thank Volker Springel for providing us with the original version of GADGET-3,
and Yehuda Hoffman for preparing the initial conditions using the Constrained 
Realizations. We are grateful to our colleagues, and especially to Kentaro Nagamine, for 
helpful discussions. I.S. acknowledges partial support by NASA and the NSF grants. 
M.T. acknowledges support by the University of Colorado ATP 
through grants from NASA and NSF. Simulations have been performed on the University
of Kentucky DLX Cluster.



\begin{thebibliography}{37}
\expandafter\ifx\csname natexlab\endcsname\relax\def\natexlab#1{#1}\fi

\bibitem[Barkana (2002)]{barkana02}
Barkana, R. 2002, New Astronomy, 7, 85

\bibitem[Barth, Martini, Nelson, \& Ho (2003)]{barth03}
Barth, A.~J., Martini, P., Nelson, C.~H., \& Ho, L.~C. 2003, \apjl, 594, L95

\bibitem[Becker et al. (2001)]{becker01}
Becker, R.~H., et al. 2001, \aj, 122, 2850

\bibitem[Berentzen \& Shlosman(2006)]{beren06}
Berentzen, I., \& Shlosman, I. 2006, \apj, 648, 807

\bibitem[Bouwens et~al.(2011)]{bouw11}
Bouwens, R.~J., et al. 2011, Nature, 469, 504

\bibitem[Bouwens et~al.(2010)]{bouw10}
Bouwens, R.~J., et al. 2010, \apj, 709, L133

\bibitem[Bouwens et~al.(2007)]{bouw07}
Bouwens, R.~J., et al. 2007, \apj, 670, 928

\bibitem[Bryan \& Norman(1998)]{bryan98}
Bryan, G.~L., \& Norman, M.L. 1998, \apj, 495, 80

\bibitem[Bunker et~al.(2004)]{bunke04}
Bunker, A.~J., Stanway, E.~R., Ellis, R.~S., \& McMahon, R.~G. 2004, \mnras, 355, 374

\bibitem[Capak et~al.(2011)]{capa11}
Capak, P.~L. et al. 2011, Nature, 470, 233

\bibitem[Cen \& McDonald (2002)]{cen02}
Cen, R., \& McDonald, P. 2002, \apj, 570, 457

\bibitem[Choi \& Nagamine (2010)]{choi10}
Choi, J., \& Nagamine, K. 2010, \mnras, 407, 1464

\bibitem[Choi \& Nagamine (2009)]{choi09}
Choi, J., \& Nagamine, K. 2009, \mnras, 393, 1595

\bibitem[Di Matteo et~al.(2005)]{dimat05}
Di Matteo, T., Springel, V., \& Hernquist, L. 2005, Nature, 433, 604

\bibitem[Dressler (1980)]{dress80}
Dressler, A. 1980, \apj, 236, 351

\bibitem[Dunkley et~al.(2009)]{dunk09}
Dunkley, J., et al. 2009, ApJS, 180, 306

\bibitem[Eisenstein \& Hut (1998)]{eisen98}
Eisenstein, D.~J., \& Hut, P. 1998, \apj, 498, 137

\bibitem[Fan et~al.(2004)]{fan04}
Fan, X., et al. 2004, \aj, 128, 515

\bibitem[Gnedin \& Kravtsov (2011)]{gned11}
Gnedin, N.~Y., \& Kravtsov, A.~K. 2011, \apj, 728, 88 

\bibitem[Gnedin et~al.(2009)]{gned09}
Gnedin, N.~Y., Tassis, K., \& Kravtsov, A.~V. 2009, \apj, 697, 55

\bibitem[Governato et~al.(2004)]{gover04}
Governato, F., et al. \apj, 607, 688

\bibitem[Graham et~al.(2001)]{graha01}
Graham, A.~W., Erwin, P., Caon, N., \& Trujillo, I. 2001, \apj, 563, L11

\bibitem[Greif et~al.(2010)]{greif10}
Greif, T.~H., Glover, S.~C.~O., Bromm, V., \& Klessen, R.~S. 2010, \apj, 716, 510

\bibitem[Harford et~al.(2008)]{harf08}
Harford, A.~G., Hamilton, A.~J.~S., \& Gnedin, N.~Y. 2008, \mnras, 389, 880

\bibitem[Heller et~al.(2007)]{hell07}
Heller, C.~H., Shlosman, I., \& Athanassoula, E. 2007, \apj, 671, 226 
 
\bibitem[Hoffman \& Ribak(1991)]{hoff91}
Hoffman, Y., \& Ribak, E. 1991, \apj, 380, L5

\bibitem[Li et~al.(2007)]{li07}
Li, Y., et al. 2007, \apj, 665, 187


\bibitem[Malhotra et~al.(2005)]{malh05}
Malhotra, S., et al. 2005, \apj, 626, 666


\bibitem[McLure et~al.(2010)]{mclure10}
McLure, R.~J., et al. 2010, \mnras, 403, 960

\bibitem[Oesch et~al.(2011)]{oesch11}
Oesch, P.~A., et al. 2011, arXiv:1105.2297

\bibitem[Oesch et~al.(2010)]{}
Oesch, P.~A., et al. 2010, \apj, 709, L21

\bibitem[Overzier et~al.(2006)]{over10}
Overzier, R.~A., Bouwens, R.~J., Illingworth, G.~D., \& Franx, M. 2006,
     \apj, 648, 5

\bibitem[Pawlik et~al.(2011)]{pawl11}
Pawlik, A.~H., Milosavljevic, M., Bromm, V. 2011, \apj, 731, 54

\bibitem[Robertson et~al.(2004)]{rober04}
Robertson, B., Yoshida, N., Springel, V., Hernquist, L. 2004, \apj, 606, 32

\bibitem[Romano-Diaz et~al.(2011)]{romano11}
Romano-Diaz, E., Shlosman, I., Trenti, M., \& Hoffman, Y. 2011, \apj, 734, in press,
           arXiv:1010.3715

\bibitem[Romano-Diaz et~al.(2009)]{romano09}
Romano-Diaz, E., Shlosman, I., Heller, C.~H., Hoffman, Y. 2009, \apj, 702, 1250

\bibitem[Romano-Diaz et~al.(2007)]{romano07}
Romano-Diaz, E., Hoffman, Y., Heller, C.~H., Faltenbacher, A., Jones, D.,
    \& Shlosman, I. 2007, \apj, 657, 56


\bibitem[Sommer-Larsen et~al.(2003)]{sommer03}
Sommer-Larsen, J., G\"otz, M., \& Portinari, L. 2003, Ap\&SS, 281, 519

\bibitem[Springel (2005)]{springel05}
Springel, V. 2005, \mnras, 364, 1101

\bibitem[Springel \& Hernquist (2005)]{spring05}
Springel, V., \& Hernquist, L. 2005, \apj, 622, L9

\bibitem[Springel et~al.(2005)]{sprin05}
Springel, V., et al. 2005, Nature, 435, 629

\bibitem[Springel \& Hernquist (2003)]{spring03}
Springel, V., \& Hernquist, L. 2003, \mnras, 339, 289

\bibitem[Springel \& Hernquist (2002)]{spring02}
Springel, V., \& Hernquist, L. 2002, \mnras, 333, 649

\bibitem[Stark et~al.(2011)]{stark11}
Stark, D.~P., Ellis, R.~S., \& Ouchi, M. 2011, \apj, 728, L2

\bibitem[Tasker et~al.(2008)]{task08}
Tasker, E.~J., \& Bryan, G.~L. 2008, \apj, 673, 810

\bibitem[Trenti et~al.(2011a)]{tren11a}
Trenti, M., et al. 2011a, ApJL, submitted

\bibitem[Trenti et~al.(2011b)]{tren11b}
Trenti, M., et al. 2011b, arXiv:1011.4075

\bibitem[Trenti et~al.(2010)]{tren10}
Trenti, M., Smith, B.~D., Hallman, E.~J., Skillman, S.~W., \& Shull, J.~M. 2010, 
    \apj, 711, 1198

\bibitem[Trenti et~al.(2008)]{tren08}
Trenti, M., Santos, M.~R., \& Stiavelli, M. 2008, \apj, 687, 1

\bibitem[Wise et~al.(2010)]{wise10}
Wise, J.~H., Turk, M.~J., Norman, M.~L., \& Abel, T. 2010, arXiv:1011.2632


\end{thebibliography}
\end{document}